\documentclass[aos]{imsart}

\RequirePackage{amsthm,amsmath,amsfonts,amssymb, bm,   tikz
}  \usetikzlibrary{positioning}

\RequirePackage[authoryear]{natbib}%% uncomment this for author-year citations
\RequirePackage[colorlinks,citecolor=blue,urlcolor=blue]{hyperref}
\RequirePackage{graphicx}

\startlocaldefs
\theoremstyle{plain}

\theoremstyle{definition}

\endlocaldefs

\begin{document}

\begin{frontmatter}
\title{Modeling Time-course Gene Expression Data through Bayesian Partition Functional Principal Component Analysis}
\runtitle{Bayesian Partition Functional Principal Component Analysis}

\begin{aug}
 
\author[A]{\fnms{Marion}~\snm{Kerioui} \ead[label=e1]{marion.kerioui@inserm.fr}\orcid{0000-0002-1694-9021}}
\author[A]{\fnms{Daniel}~\snm{Temko} \orcid{0000-0002-0165-292X
}}
\author[B]{\fnms{Shahin}~\snm{Tavakoli}\orcid{0000-0002-6323-6227}}
\author[A]{\fnms{Hélène}~\snm{Ruffieux}\ead[label=e3]{helene.ruffieux@mrc-bsu.cam.ac.uk}\orcid{0000-0002-7113-2540}}
 
\address[A]{MRC Biostatistics Unit, Cambridge University, United Kingdom\printead[presep={ ,\ }]{e1,e3}}

\address[B]{RISIS, GSEM, University of Geneva, Geneva, Switzerland}
\end{aug}

\begin{abstract}
 High-dimensional biomarkers such as gene expression levels are now routinely measured over time, allowing biological processes to be studied dynamically rather than through cross-sectional snapshots. However, existing methods do not adequately address the central applied challenges posed by such data: simultaneously reducing dimensionality, quantifying inter-individual variability and uncovering temporal structure shared across biomarkers.
   We introduce Partition Functional Principal Component Analysis (PFPCA), a Bayesian model that jointly learns shared temporal patterns and clusters variables according to their latent dynamics. PFPCA combines a mixture model
with multivariate functional principal component analysis performed within each group.  
     We develop a scalable mean-field variational algorithm for joint inference of functional principal component loadings, individual-level scores, group assignments and partition sizes.
       Simulations show clear gains from joint inference: PFPCA recovers both the partition and the latent functional structure more accurately than a two-step baseline. In the most challenging settings, PFPCA retrieves the true partition in 27\% of replicates compared with 1\% for the two-step baseline. 
      Applied to longitudinal gene-expression data from individuals experimentally infected with H3N2 influenza virus, PFPCA identifies groups of genes with coordinated activation patterns and reveals temporal signatures associated with immune-response dynamics and symptom status. 
\end{abstract}

% \begin{keyword}[class=MSC]
% \kwdgroup[type=primary]{\kwd{00X00}
% \kwd{00X00}}
% \kwdgroup[type=secondary]{\kwd{00X00}}
% \end{keyword}

\begin{keyword}
\kwd{High-dimensional data}
\kwd{Longitudinal data}
\kwd{Mixture model}
\kwd{Variational inference}
\end{keyword}

\end{frontmatter}
%%%%%%%%%%%%%%%%%%%%%%%%%%%%%%%%%%%%%%%%%%%%%%
%% Please use \tableofcontents for articles %%
%% with 50 pages and more                   %%
%%%%%%%%%%%%%%%%%%%%%%%%%%%%%%%%%%%%%%%%%%%%%%
%\tableofcontents

\section{Introduction}\label{sec-intro}

In biomedical studies, an increasing number of biomarkers are measured repeatedly over time, enabling the study of biological processes underlying disease progression or response to treatment. Such longitudinal measurements offer the possibility to move beyond static
snapshots and to characterize how coordinated biological systems evolve dynamically
across individuals. In particular, they raise fundamental biological questions, such as
which subsets of biomarkers exhibit coherent temporal behavior, how these dynamics vary
across subjects and how they relate to clinically relevant outcomes. Despite the richness
of these data, addressing such questions remains challenging in high-dimensional settings
and key aspects of coordinated longitudinal behavior remain difficult to extract in a
principled and interpretable way. 

In this paper, we consider a longitudinal gene expression study of healthy individuals
experimentally inoculated with the H3N2 influenza virus \citep{chen_predicting_2011}, a setting that
exemplifies the challenges posed by high-dimensional longitudinal biomarker data typical of modern biomedical studies. The scale and complexity of such data exceed the scope of standard longitudinal and functional data methods. 
In particular, extracting meaningful biological insights from high-dimensional longitudinal biomarkers requires methods that simultaneously address two key aspects: first, reducing dimensionality by identifying subsets of genes that covary over time, thereby partitioning genes into biologically relevant groups and, second, characterizing the temporal evolution of the level of expression of each gene in each individual. 
To our knowledge, no existing approach addresses these objectives jointly
in a single, principled framework, motivating the methodological developments introduced
in this paper.

 While there is a broad literature focusing on model-based clustering of \emph{individuals} based on time-dependent variables \citep[e.g.,][]{lu_clustering_2024, jacques_model-based_2014, proust-lima_estimation_2017, bouveyron_model-based_2011}, comparatively less attention has been devoted to methods for clustering a large number of 
 time-dependent \emph{variables}. A few Bayesian hierarchical models \citep{heard_quantitative_2006, ma_bayesian_2008, angelini_clustering_2012} do cluster variables according to their temporal patterns, but these approaches are designed for data arising from a single individual.  As a result, the identified clusters may be driven by idiosyncratic features of one individual, and they do not extend naturally to settings where variables are repeatedly measured across several individuals. A more general approach is therefore needed; one that models between-individual variability explicitly and yields variable clusters that are robust across subjects.

Functional Principal Component Analysis (FPCA) offers a powerful framework for uncovering principal modes of variation from functional data. FPCA extends classical PCA to functional data by representing each trajectory, via the Karhunen--Loève expansion, as a mean function plus a finite number of orthogonal eigenfunctions capturing the dominant modes of temporal variation \citep{goldsmith_generalized_2015}. Individual-specific scores associated with these eigenfunctions provide a low-dimensional summary of longitudinal dynamics. Multivariate FPCA (MFPCA), which is based on the multivariate extension of the Karhunen--Loève theorem \citep{everitt_functional_2005, happ_multivariate_2018}, further allows coordinated analysis of multiple functional variables by assuming a shared underlying latent process. However, this assumption becomes unrealistic in high-dimensional settings, where only subsets of variables are expected to share common temporal dynamics. Applying MFPCA globally in such contexts can obscure structure rather than reveal it. This limitation highlights the need for an approach that adaptively identifies groups of variables for which shared latent dynamics are plausible.

 To address this need, we introduce Partition Functional Principal Component Analysis (PFPCA), a Bayesian framework that simultaneously partitions variables into groups driven by shared latent temporal processes and performs MFPCA within each group. 
 
    Building on recent advances in Bayesian FPCA \citep{nolan_bayesian_2025,nolan_efficient_2025}, we develop a variational inference algorithm that jointly estimates all parameters and is specifically designed for computational scalability in high dimension. 
    By jointly modeling genes within groups, PFPCA enables principled  information sharing across variables, leading to improved parameter estimation and more reliable inference on the underlying latent functions. To further enhance exploration of multimodal posterior distributions, we combine the variational algorithm with a simulated annealing procedure \citep{kirkpatrick_optimization_1983}.
 
This article is organized as follows. Section~\ref{sec:meth} introduces the Bayesian partition FPCA model after a brief reminder on MFPCA. Section~\ref{sec:infe} details the inference strategy. Section~\ref{sec:simu} presents a comprehensive simulation study assessing estimation accuracy and comparing performance with alternative approaches. Finally, Section~\ref{sec:appli} analyzes the motivating dataset to characterize coordinated gene dynamics in response to infection.  An R package implementing PFPCA is available on Github at \url{https://github.com/keriouim/partFPCA}.

\section{Methods}
\label{sec:meth}
In this section, we provide a brief reminder of multivariate FPCA (MFPCA) before introducing our proposed partition FPCA (PFPCA) model.
\subsection{Multivariate Functional Principal Component Analysis}

Consider a series of $p$ random function $y^{(j)}: [0,1] \rightarrow \mathbb{R}$, $j = 1, \ldots, p$, and write the corresponding vector of random functions $y=\left[y^{(1)}\, \cdots \, y^{(p)}\right]^{\intercal}$ which we assume is an element of the Hilbert space $\mathcal{H} = \left(L^2([0,1])\right)^p$ equipped with the inner product: 
\begin{equation*}
\langle f, g\rangle_{\mathcal{H}}= \sum_{j=1}^p \int_0^1 f^{(j)}(t) g^{(j)}(t) \mathrm{d}t, ~~~f, g\in \mathcal{H}.
\end{equation*}
 
Given the existence of continuous mean and covariance functions, the multivariate version of the Karhunen–Loève theorem provides the basis for the MFPCA decomposition~\citep{happ_multivariate_2018}.  Given random realizations $y_1, \ldots, y_N$ of $y$   
there exists a complete orthonormal basis of functions $\psi_l \in \mathcal{H},$ for $l \in \mathbb{N}$, such that 
\begin{equation}\label{eq:KLdecomposition}
    y_i = \mu+ \sum_{l=1}^{\infty} \zeta_{i,l} \psi_l, ~~~ i=1,\ldots,N,  
\end{equation} 
where $\mu =\left[ \mathbb{E}\left(y_i^{(1)}\right)\, \cdots\,\mathbb{E}\left(y_i^{(p)}\right) \right]^\intercal$, and $\zeta_{i,l}= \langle y_i-\mu , \psi_l \rangle_{\mathcal{H}}$ are the principal component scores. These scores are independent across $i$ and uncorrelated across $l$, with $\mathbb{E}(\zeta_{i,l})=0$ and $\mathbb{V}\text{ar}(\zeta_{i,l})=\gamma_l$, where $\gamma_l$ is the $l^{\text{th}}$ largest eigenvalue of the multivariate covariance operator. We further have $\gamma_l\rightarrow 0$ as $l \rightarrow \infty,$ which justifies the approximation  
\begin{equation}\label{eq:KLtruncated} 
    y_i \approx \mu+ \sum_{l=1}^{L} \zeta_{i,l} \psi_l,  ~~~ i=1,\ldots,N,  
\end{equation} 
for $L$ sufficiently large.  

In practice, functional data are obtained at discrete time points with an error, that we assume to be Gaussian. Hence for the $i^{\text{th}}$ realization (which may correspond to an ``individual'') of the $j^{\text{th}}$ random function (``variable'' or ``biomarker'' in a biomedical setting), we observe the response vector $$\bm{y}_i^{(j)}= y_i^{(j)}(\bm{t}_i^{(j)})+\bm{\varepsilon}_{i}^{(j)},$$ where $\bm{t}_i^{(j)}= \left(t_{i,1}^{(j)}, \dots, t_{i,n_{i}^{(j)}}^{(j)}\right)^{\intercal}$ denotes the grid of $n_{i}^{(j)}$ measurement times for the corresponding individual and variable, and $$\bm{\varepsilon}_{i}^{(j)} \sim\mathcal{N}\left(\bm{0}_{n_{i}^{(j)}}, \left(\sigma_{\varepsilon}^{(j)}\right)^2 \bm{I}_{n_{i}^{(j)}} \right).$$ For notational convenience --- although the method supports different time grids across individuals and across variables --- we now assume that the time grid is the same for all individuals and all variables, i.e., $\bm{t}_i^{(j)}=\bm{t}$ and $n_{i}^{(j)}=n$, for $i = 1, \dots, N$ and $j = 1,\dots, p.$
Letting $\bm{\mu}^{(j)}=\mu^{(j)}(\bm{t})$ and $\bm{\psi}_{l}^{(j)}= \psi_{l}^{(j)}(\bm{t})$ be the vectors corresponding to the mean function and eigenfunctions respectively, assessed at these time points, the corresponding statistical model can thus be written as:
\begin{align}\bm{y}_i^{(j)}=\bm{\mu}^{(j)}+\sum_{l=1}^L\zeta_{i,l}\bm{\psi}_{l}^{(j)}+\bm{\varepsilon}_{i}^{(j)}.
\end{align}

The decomposition in Equation \eqref{eq:KLtruncated} is unique up to a change of sign of the scores and eigenfunctions, provided all eigenfunctions are distinct \citep[see, e.g., Theorem 3.1, ][]{nolan_efficient_2025}. This property allows for an initial estimation of the decomposition without imposing any constraint on the scores and latent functions, thereby avoiding the computationally intensive task of estimating the covariance function \citep{happ_multivariate_2018}. For example, \citet{nolan_efficient_2025} proposed a Bayesian inference framework for the MFPCA, where the basis of functions is orthonormalized and the scores are decorrelated post inference, to retrieve the unique decomposition. 

Intuitively, the scores control the magnitude of the deviation from the mean function for a specific realization, and sharing scores across variables accounts for the shared latent process driving the curves' evolution. However, when the number of variables becomes larger, it may be unreasonable to assume that all variables share the same set of scores, as groups of variables may be driven by different latent dynamics. In the next section, we introduce our proposed model, that accounts for this group structure. 
\subsection{Bayesian Partition Functional Principal Component Analysis}\label{sec_model}

We propose a new hierarchical extension of multivariate functional principal component analysis, which we refer to as \emph{Partition FPCA} (PFPCA). The PFPCA model incorporates  an additional level of model hierarchy that groups variables whose curves reflect the influence of a shared latent process.

Consider $Q$ groups and let $z_j\in \{1, \dots,Q\}$ denote a latent class variable, i.e., $z_j=q$ if the $j^{\rm{th}}$ variable belongs to group $q$. Conditionally on $z_j=q$, the model assumes that the vector of observations $\bm{y}_{i}^{(j)}$ follows a MFPCA model where the vector of $L$ scores, $\bm{\zeta}_i^{(q)}= \left(\zeta_{i,1}^{(q)}, \dots ,\zeta_{i,L}^{(q)}\right)^\intercal$, is shared across the variables belonging to group $q$:
 \begin{align*}
    \bm{y}_i^{(j)}\mid z_j=q, \bm{\mu}^{(j)}, \bm{\zeta}_{i}^{(q)}, \bm{\Psi}^{(j,q)}, \left(\sigma_{\varepsilon}^{(j,q)}\right)^2\sim \mathcal{N}\left(\bm{\mu}^{(j)}+\sum_{l=1}^L\zeta_{i,l}^{(q)}\bm{\psi}_{l}^{(j,q)}, \left(\sigma_{\varepsilon}^{(j,q)}\right)^2\times \bm{I}_{n}\right),
\end{align*}
where $\bm{\mu}^{(j)}= \mu^{(j)}(\bm{t}) $ is the mean function vector for variable $j$, assessed at the vector of time points $\bm{t}$ and $\bm{\psi}_l^{(j,q)}=\psi_l^{(j,q)}(\bm{t})$ is the $l^\text{th}$ eigenfunction vector for variable $j$ (belonging to group $q$), assessed at this same vector of time points, with  $\bm{\Psi}^{(j,q)}=\left(\bm{\psi}_1^{(j,q)},\cdots, \bm{\psi}_L^{(j,q)} \right)^{\intercal} $ the matrix summarizing the $L$ eigenfunction vectors, and $\sigma_{\varepsilon}^{(j,q)2}$ the variance of the associated residual error. We further place a categorical prior on the latent variable $z_j$: \begin{align}
    z_j \mid \bm \theta \sim \mathcal{C}at(Q, \bm{\theta}), \hspace{0.5cm} \bm{\theta} \sim \mathcal{D}ir(\bm{\alpha}), \hspace{0.5cm} & ~ j=1,\dots,p, 
\end{align}
where the mixture proportion $\bm{\theta}$ follows a Dirichlet distribution with concentration parameter $\bm{\alpha}= (\alpha_1, \dots, \alpha_Q)^\intercal$. Here, we set $\alpha_q=1/Q$ for $q=1,\dots, Q$; this is a reasonable choice in absence of prior belief on the partition \citep{ickstadt_toward_2018, rousseau_asymptotic_2011}. Sensitivity to hyperparameter $\alpha_q$ will be discussed further in Section \ref{sec_choices}. 
Writing $\bm{\sigma}^{(j)2}_\varepsilon=\left( \sigma_\varepsilon^{(j,1)2}, \ldots, \sigma_\varepsilon^{(j,Q)2} \right)^{\intercal}$ the vector of error variances, we can therefore re-write the likelihood as a mixture of MFPCA models:
\begin{align*}
    \bm{y}_i^{(j)}\mid\bm{\theta}, \bm{\sigma}_{\varepsilon}^{(j)2},\bm{\mu}^{(j)}, \left( \bm{\Psi}^{(j,q)},\bm{\zeta_i}^{(q)}\right)_{q=1,\dots,Q}\sim \sum_{q=1}^Q \theta_{q} \times  \mathcal{N}\left(\bm{\mu}^{(j)}+\sum_{l=1}^L\zeta_{i,l}^{(q)}\bm{\psi}_{l}^{(j,q)} , \left(\sigma_{\varepsilon}^{(j,q)}\right)^2\times \bm{I}_{n}\right).
\end{align*}

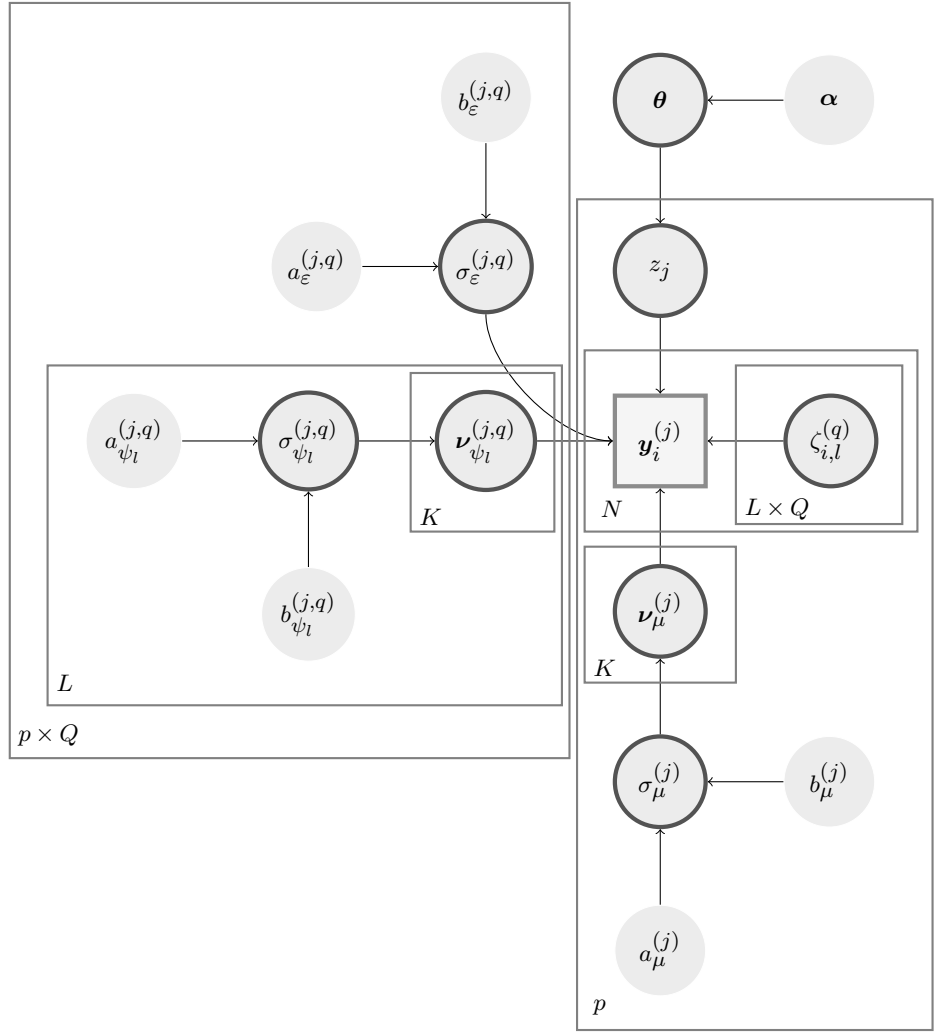
\begin{figure}[h]
    \centering
    \begin{tikzpicture}[
roundnode/.style={circle, draw=darkgray!90, fill=darkgray!10,  ultra thick, minimum size=12mm},
roundnode2/.style={circle, draw=darkgray!0, fill=darkgray!10, thin, minimum size=12mm},
squarednode/.style={rectangle, draw=darkgray!60, fill=darkgray!5, ultra thick, minimum size=12mm},
]

%Nodes
\node[squarednode]      (maintopic)                              {$\bm{y}_i^{(j)}$};
\node[roundnode]        (latent)       [above=of maintopic] {$z_j$};
\node[roundnode]        (theta)       [above=of latent] {$\bm{\theta}$};
\node[roundnode2]        (alpha)       [right=of theta] {$\bm{\alpha}$};
\node[roundnode]      (scores)       [right=of maintopic] {$\zeta_{i,l}^{(q)}$};

 \node[roundnode]      (eigennu)       [left=of maintopic] {$\bm{\nu}_{\psi_l}^{(j,q)}$};
  \node[roundnode]      (meannu)       [below=of maintopic] {$\bm{\nu}_{\mu}^{(j)}$};
    \node[roundnode]      (sigmamu)       [below=of meannu] {$\sigma_{\mu}^{(j)}$};
     \node[roundnode2]      (amu)       [below=of sigmamu] {$a_{\mu}^{(j)}$};
   \node[roundnode2]      (bmu)       [right=of sigmamu] {$b_{\mu}^{(j)}$};
 \node[roundnode]      (sigmaeps)       [above=of eigennu] {$\sigma_{\varepsilon}^{(j,q)}$};
  \node[roundnode]      (sigmapsi)       [left=of eigennu] {$\sigma_{\psi_l}^{(j,q)}$};
   \node[roundnode2]      (apsi)       [left=of sigmapsi] {$a_{\psi_l}^{(j,q)}$};
   \node[roundnode2]      (bpsi)       [below=of sigmapsi] {$b_{\psi_l}^{(j,q)}$};
  \node[roundnode2]      (aeps)       [left=of sigmaeps] {$a_{\varepsilon}^{(j,q)}$};
   \node[roundnode2]      (beps)       [above=of sigmaeps] {$b_{\varepsilon}^{(j,q)}$};

%Lines
\draw[->] (aeps.east) -- (sigmaeps.west);
\draw[->] (beps.south) -- (sigmaeps.north);
\draw[->] (apsi.east) -- (sigmapsi.west);
\draw[->] (bpsi.north) -- (sigmapsi.south);
\draw[->] (amu.north) -- (sigmamu.south);
\draw[->] (bmu.west) -- (sigmamu.east);
\draw[->] (latent.south) -- (maintopic.north);
\draw[->] (theta.south) -- (latent.north);
\draw[->] (meannu.north) -- (maintopic.south);
\draw[->] (sigmamu.north) -- (meannu.south);
\draw[->] (sigmaeps.south) .. controls +(down:7mm) and +(left:7mm) .. (maintopic.west);
\draw[<-] (theta.east) -- (alpha.west);
\draw[<-] (maintopic.east) -- (scores.west);
\draw[->] (eigennu.east) -- (maintopic.west);
\draw[->] (sigmapsi.east) -- (eigennu.west);
%boxing
	\draw [color=gray,thick](1,-1.1) rectangle (3.2,1);
    	\node at (1,-0.9) [above=5mm, right=0mm] {\textsc{$L\times Q$}};
        
        	\draw [color=gray,thick](-1,-1.2) rectangle (3.4,1.2);
            \node at (-0.9,-0.9) [above=5mm, right=0mm] {\textsc{$N$}};
            
\draw [color=gray,thick](-1.1,-7.8) rectangle (3.6,3.2);
\node at (-1,-7.5) [above=5mm, right=0mm] {\textsc{$p$}};

\draw [color=gray,thick](-1,-3.2) rectangle (1,-1.4);
\node at (-1,-3) [above=5mm, right=0mm] {\textsc{$K$}};

\draw [color=gray,thick](-1.4,0.9) rectangle (-3.3,-1.2);
\node at (-3.3,-1) [above=5mm, right=0mm] {\textsc{$K$}};

\draw [color=gray,thick](-1.3,1) rectangle (-8.1,-3.5);
\node at (-8.1,-3.2) [above=5mm, right=0mm] {\textsc{$L$}};

\draw [color=gray,thick](-1.2,5.8) rectangle (-8.6,-4.2);
\node at (-8.6,-3.9) [above=5mm, right=0mm] {\textsc{$p\times Q$}};

\end{tikzpicture}

     \caption{\textbf{Directed Acyclic graph (DAG) of PFPCA Bayesian model}. Square indicates observed variables and circles corresponds to estimated (circled) or fixed parameters.  }
    \label{fig:dag}
\end{figure}

We use a semi-parametric regression model to describe the mean function and eigenfunctions with a basis of $K^{(j)}$ spline functions, and 
 we denote $\bm{\nu}_{\mu}^{(j)}$ the vector of $K^{(j)}$ spline coefficients specific to the mean function and $\bm{\nu}_{\psi}^{(j,q)}$ the matrix of dimension $ L \times K^{(j)} $ summarizing the vectors of spline coefficients used to describe the eigenfunctions $\bm{\nu}_{\psi}^{(j,q)} = ~\left(\bm{\nu}_{\psi_1}^{(j,q)}, \cdots, \bm{\nu}_{\psi_L}^{(j,q)} \right)^{\intercal}$. Although other basis of functions could be considered, here we use O'Sullivan penalised splines as described in \citep{wand_semiparametric_2008} to prevent overfitting (see Supplementary Material for detailed definition of the semi-parametric model and choice of the number of spline functions). We place Gaussian distributions on each of these vectors with diagonal variance-covariance matrices with diagonal elements $\left(\sigma_\mu^{(j)}\right)^2,\left(\sigma_{\psi_1}^{(j,q)}\right)^2,\dots, \left(\sigma_{\psi_L}^{(j,q)}\right)^2 $ respectively. Finally, we place Gaussian prior distributions on the scores and Gamma prior distributions on the precision terms:
 \begin{align*}
  & \bm{\zeta}_i^{(q)} \sim \mathcal{N}\left( \bm{0}_L, \bm{I}_L \right), & i = 1, \dots, N,  ~ q= 1, \dots, Q,\\
   &\left(\sigma_{\varepsilon}^{(j,q)}\right)^{-2}\mid a_{\varepsilon}^{(j,q)}, b_{\varepsilon}^{(j,q)} \sim \textnormal{Gamma}(a_\varepsilon^{(j,q)}, b_\varepsilon^{(j,q)} ), &~ j=1,\dots,p,
\end{align*}
 with shape and scale parameters $a_\varepsilon^{(j,q)}>0$ and  $b_\varepsilon^{(j,q)}>0$, and similarly for the precision terms $\left(\sigma_{\mu}^{(j)}\right)^{-2}$ and $\left(\sigma_{\psi_l}^{(j,q)}\right)^{-2}$. In absence of prior knowledge, it is reasonable to set $a_\mu^{(j)}=b_\mu^{(j)}=1$ for $j=1, \dots, p$,
 $a_{\psi_l}^{(j,q)}=l,$ and $b_{\psi_l}^{(j,q)}=1$ for $l=1,\dots,L$ and $q =1, \dots, Q$, therefore encouraging decreasing proportion of variance explained by the subsequent components, as assumed by the model. By default, we set $a_{\varepsilon}^{(j,q)}=b_{\epsilon}^{(j,q)}=10$.  Figure \ref{fig:dag} displays the Directed Acyclic Graph of the PFPCA model. In the following, we denote $\bm{\tau}$ the vector of all model parameters. 

\section{Inference}\label{sec:infe}

In this section, we present an annealed variational algorithm for efficient inference of our PFPCA model parameters in high dimension.

\subsection{Mean-field variational Bayes algorithm}

In this section, we denote the data for the response curve of each individual by $\bm{y}_i=\left(\bm{y}_i^{(1) \intercal}, \dots, \bm{y}_i^{(p) \intercal }\right)^{\intercal} $ and the full data as $\bm{y}=\left(\bm{y}_1^{ \intercal }, \dots, \bm{y}_N^{\intercal }\right)^{\intercal}$. We aim to estimate the posterior density $p\left(\bm{\tau} \mid \bm{y} \right)$. Inference of intractable posterior commonly relies on Markov chain Monte Carlo (MCMC) algorithms, which can be computationally costly, as in our high-dimensional longitudinal setting. We therefore rely on a faster deterministic approximation strategy, through a novel variational algorithm tailored to our model. Variational inference involves choosing a family of density functions to approximate the posterior distribution. It provides tractable approximations by selecting a density $q(\bm{\tau} )$, within this family, that minimizes the Kullback–Leibler divergence to $p\left(\bm{\tau} \mid \bm{y} \right)$:
\begin{equation}\label{eq:KL}
    \text{KL}(q \| p)= - \int q(\bm{\tau}) \log\left\{ \frac{p(\bm{\tau}\mid \bm{y})}{q(\bm{\tau})} \right\}d\bm{\tau}.
\end{equation}
Since 
$$\text{KL}\left(q \| p \right) =   
\log p(\bm{y})  + \int q(\bm{\tau}) \log q(\bm{\tau})d\bm{\tau}  - \int q(\bm{\tau}) \log p(\bm y, \bm{\tau})d\bm{\tau}, 
$$
minimizing \eqref{eq:KL} is equivalent to maximizing the lower bound on the marginal log-likelihood \citep{blei_variational_2017}, called the Evidence Lower BOund (ELBO): 
\begin{align*}
    \text{ELBO}(q)= \int q(\bm{\tau}) \log \left\{ \frac{p(\bm{y}, \bm{\tau})}{q(\bm{\tau})}\right\} d\bm{\tau}. 
\end{align*}

We use a mean-field variational family, which factorizes the approximation $q$ according to the components of the parameter vectors $\bm \tau$. This family entails a trade-off between faithful approximation and computational tractability. Specifically, we assume the following factorization:
\begin{equation}\label{eq:decompMFVB}
\begin{split}
    q(\bm \tau)& = \prod_{j=1}^p \left[ q\left( z_j   \right) q\left(\bm{\nu}_{\mu}^{(j)}\right)q\left( \sigma_{\mu}^{(j)}   \right)  \prod_{q=1}^Q \left\{ q\left(\bm{\nu}_{\psi}^{(j,q)}\right)  q\left( \sigma_{\varepsilon}^{(j,q)} \right)    \prod_{l=1}^L q\left( \sigma_{\psi_l}^{(j,q)}  \right) \right\}  \right]\\
    & \times q\left(  \bm{\theta} \right)\prod_{q=1}^Q   \prod_{i=1}^N q\left(\bm{\zeta}_i^{(q)}\right).\\       
\end{split}
\end{equation}

 \citet{nolan_efficient_2025} have used a similar mean-field approximation for variational inference in the context of the MFPCA model. Following previous work \citep{nolan_bayesian_2025}, it assumes posterior independence of the mean function and eigenfunctions (i.e., the population parameters) from the individual scores. Some factorizations are induced by conditional independence \citep{bishop_pattern_2006}, that can be directly deducted from our DAG in Figure \ref{fig:dag}, e.g., posterior independence between $\bm{\theta}$ and all other population parameters. This decomposition also relies on asymptotic independence between regression coefficients and variance parameters \citep{menictas_variational_2013}, e.g., posterior independence between $\bm{\nu}_\mu^{(j)}$ and $\sigma_\mu^{(j)}.$ 
  Under this assumption, the parameters for each of the $q-$densities can be obtained iteratively via a coordinate ascent algorithm. Thanks to the factorization in Equation \eqref{eq:decompMFVB}, all updates have closed-form expressions (as detailed in section 2 in Supplementary Material), making the resulting procedure computationally efficient.
  
  As we conduct inference using a variational algorithm that does not enforce orthonormal constraints on the FPCA decomposition, we subsequently apply a post-inference orthogonalisation procedure within each group of variables, as described in \citet{nolan_efficient_2025}. This procedure uses the variational posterior mean of the scores and the variational posterior mean of eigenfunctions at each time. In the following, the quantities $\widehat{\zeta}_{il}^{(q)}$ and $\widehat{\sigma}^{(q)}_{\zeta,l}$ refer to the posterior estimates of the scores and their variance obtained after post-orthonormalization. 

  \subsection{Simulated annealing procedure}
  
  As the posterior distribution of our proposed model is highly multimodal and variational inference only guarantees convergence to a local mode, we couple our algorithm with a simulated annealing procedure \citep{ueda_deterministic_1998} to enhance the exploration of the posterior parameter space. Annealing introduces a so-called ``temperature'' parameter $T>0$, which smooths out the posterior density during the first iterations of the algorithm. This temperature is gradually reduced, following a geometric cooling schedule \citep{kirkpatrick_optimization_1983}, until it reaches $T = 1$ corresponding to the original optimization problem, after which the standard variational algorithm is run until convergence \citep{ruffieux_global-local_2020}. As is intuitive, the temperature parameter acts as a multiplier on the entropy term in the ELBO: $$\text{ELBO}_{T}(q)=\int q(\bm{\tau}) \log(p(y, \bm{\tau}))  d\bm{\tau} - T\int q(\bm{\tau}) \log(q(\bm{\tau})) d\bm{\tau}.$$ When the temperature is high, it places greater emphasis on the entropy, leading to more diffuse (i.e., less certain) variational distributions in the early stages of inference.  For our model, the annealed approximate posterior updates and ELBO derivations can be found in Section 2 in Supplementary Material. In the following, we will refer to this algorithm as the annealed mean-field variational Bayes (AMFVB) algorithm.

There are two hyperparameters to choose when performing simulated annealing, the initial temperature $T$ and the number of annealed iterations. There is no consensus in the literature on optimal values of these parameters, and the number of iterations is generally chosen empirically to improve performance of the algorithm while avoiding an unnecessary increase in computational burden. Here, $T$ can not be larger than $K/2$ or $N/2$ to ensure strict positivity of the shape parameters in the inverse-gamma distributions. In this context, we found limited impact of this latter parameter on the performance of the algorithm, while a low number of annealed iterations leads to poorer partitioning. In practice, we recommend using $T=2$ and $100$ iterations of the annealed algorithm, which is the default setting in our algorithm. Our R package \texttt{partFPCA} is available at \url{https://github.com/keriouim/partFPCA}.

\subsection{Learning the number of groups and FPCA components}\label{sec_choices}
The selection of the number of components within the FPCA expansion of each group is conducted as follows. We perform inference assuming a ``conservatively large'' number of components $L^{(q)} \equiv L$ for all the groups, say $L = 10$ or $20$. We then select the effective number of components necessary to account for the variation in each group by estimating the proportion of variance explained by the first few components. Recall that, according to the Karhunen–Loève theorem, the variances of the scores correspond to the eigenvalues of the covariance operator. Therefore, letting $\widehat{\sigma}_{\zeta, l}^{(q)2}$ denote the empirical variance of the estimated scores $\left(\widehat{\zeta}_{il}^{(q)}\right)_{i=1,\dots,N}$, obtained after orthonormalization, the proportion of variance explained by the $l^{th}$ component can be estimated as: $$\widehat{\text{PV}}^{(q)}_l=\frac{\widehat{\sigma}_{\zeta, l}^{(q)2}}{\sum_{l=1}^L \widehat{\sigma}_{\zeta, l}^{(q)2}}.$$ The cumulative proportion of variance explained by the first components is then $\widehat{\text{CPV}}^{(q)}_l=\sum_{m=1}^l\widehat{\text{PV}}^{(q)}_m,$ for $l= 1, \dots, L.$ For each group $q\in\{1, \ldots Q\}$, we select the minimum number of components for which the model explains 95\% of the overall variability, i.e., the smallest $l$ verifying $\widehat{\text{CPV}}^{(q)}_l\geq 0.95$.

Similarly, we set the number of mixture components $Q$ to a ``large'' value, a situation commonly referred to as overfitted mixture model \citep{papastamoulis_overfitting_2018}. We then determine the optimal number of groups by counting the number $Q^*$ of non-empty groups, with a group being considered empty when no variable is assigned to this group \textit{a posteriori}. Both $Q$ and the concentration parameter $\bm{\alpha}$ govern the number of non-empty groups \textit{a priori}, as illustrated in \cite{fruhwirth-schnatter_here_2019}. In absence of prior belief on the partition, it is reasonable to consider a symmetric Dirichlet distribution, i.e., $\alpha_q=\alpha$ for all $q \in \{1, \dots, Q\}. $ Choosing $\alpha<1$ encourages sparsity in the number of non-empty groups and, under certain assumptions, guarantees asymptotic convergence to the true number of groups \citep{rousseau_asymptotic_2011}. We recommend setting $\alpha$ to $Q^{-1}$, which is supported by both the literature \citep[e.g.,][]{ickstadt_toward_2018} and sensitivity analyses conducted in the context of our model (see Supplementary Material).

\section{Simulation studies}\label{sec:simu}

This section presents a series of numerical experiments designed to assess the statistical and computational performance of our approach, and compare it with established approaches.

\subsection{Data generation and performance metrics}\label{sec_settings}
For each simulation study, we generate $100$ datasets as follows.
We consider a partition $\mathcal{G}=\left\{ G_q \right\}_{ q = 1,\dots,Q}$ of $\{1, \dots,p\}$, where $G_q$ gathers the indices of the functional variables in group $q$, and we let $|G_q|$ denote the size of this group. Then for each group $q$, we simulate an MFPCA expansion by sampling randomly, for each variable within the group, $L$ eigenfunctions from 11 periodic B-splines of order 4 transformed using a Gram-Schmidt orthogonalisation procedure (using the R package \texttt{splinets}), and rescaled by $\sqrt{|G_q|}$ to ensure orthonormality in the multivariate Hilbert space within each group. We further sample the vector of scores $\bm{\zeta}_i^{(q)}$, for $i = 1, \ldots, N$, from a Gaussian distribution $\mathcal{N}(\bm{0}_L, \bm\Sigma_\zeta^{(q)})$ with $\bm\Sigma_\zeta^{(q)}= \text{diag}\left(|G_q| \times v_l\right)_{l=1,\dots,L},$ where $v_l>0$ is decreasing with $l$ and depends on the simulation scenario. This ensures that the contribution of the principal components and the error to the total variance remains similar across groups of varying sizes. We simulate the mean function of  each variable $j \in \{1,\dots, p\}$ as $\mu^{(j)}(t)= (-1)^j\times \sin\left\{2\pi + (j \bmod 5) \times t\right\}$, where $\bmod$ is the remainder of the Euclidean division, and we set the variance of the measurement error, $\left(\sigma_{\varepsilon}^{(j)}\right)^2$, to $1$. We sample the number of time observations for each individual from a Poisson distribution with parameter $n_{\lambda}$, whose specification depends on the scenario, and the observation times from a uniform distribution over interval $[0,1]$.

We assess estimation accuracy using the root mean square error (RMSE) for the scores, and the integrated squared error (ISE) for the mean function, eigenfunctions and individual curves, defined as $$\text{ISE}(f, \widehat{f})=\int_0^1 \left(f(x)-\widehat{f}(x)\right)^2 \mathrm{d}x,$$ with $f(\cdot)$ the function used for simulating the data and $\widehat{f}(\cdot)$ its corresponding posterior estimate.   
Finally, we use the adjusted rand index (ARI) as a measure of the quality of the partitioning \citep{hubert_comparing_1985}. The ARI provides a measure of similarity between two partitions --- here, the ``true'' partition used to generate the data and the partition recovered from the PFPCA estimation. The ARI ranges between $0$ and $1$, with $1$ indicating identical partitions. 

Unless otherwise stated, we estimate the parameters of the PFPCA model with $Q=10$ mixture components and $L=10$ FPCA components initially, and we learn the optimal number of groups and principal components from the data, as described in Section~\ref{sec_choices}. Posterior estimates are obtained using our proposed AMFVB algorithm, running $100$ annealed iterations of the algorithm with an initial annealing temperature set to $T=2$ and then until convergence, as assessed based on the relative changes of the ELBO with tolerance $10^{-5}$.

\subsection{Accuracy of the variational algorithm}

In this section, we compare the PFPCA model's parameter estimates obtained using our proposed mean-field variational (AMFVB) algorithm, with   
an MCMC Hamiltonian Monte Carlo (HMC) algorithm implemented for the same model using the R package \texttt{rstan}. Due to the high computational cost associated with sampling-based algorithms, we restrict the dataset size for this comparison to $p=6$ variables evenly split between $Q=2$ groups ($|G_1|=|G_2|$), each observed longitudinally for $N=200$ individuals, at $n_{\lambda}=20$ timepoints on average. We simulate $L=2$ eigenfunctions with variance of the scores $v_l = l^{-2}$ for $l =1, 2$. 
 To make HMC inference computationally feasible, we assume the number of eigenfunctions and of groups to be known, i.e., we take $L=Q=2$ for estimation; in all subsequent sections, these quantities will be learnt from the data, as outlined in Section~\ref{sec_choices}.  Finally, to ensure a fair runtime comparison between the variational and MCMC inference methods, we run both algorithms serially. In this setting, the two groups of variables are exchangeable, so we only display results for one of them for the sake of readability.  
 
\begin{figure}[h]
    \centering

    \includegraphics[scale=0.4]{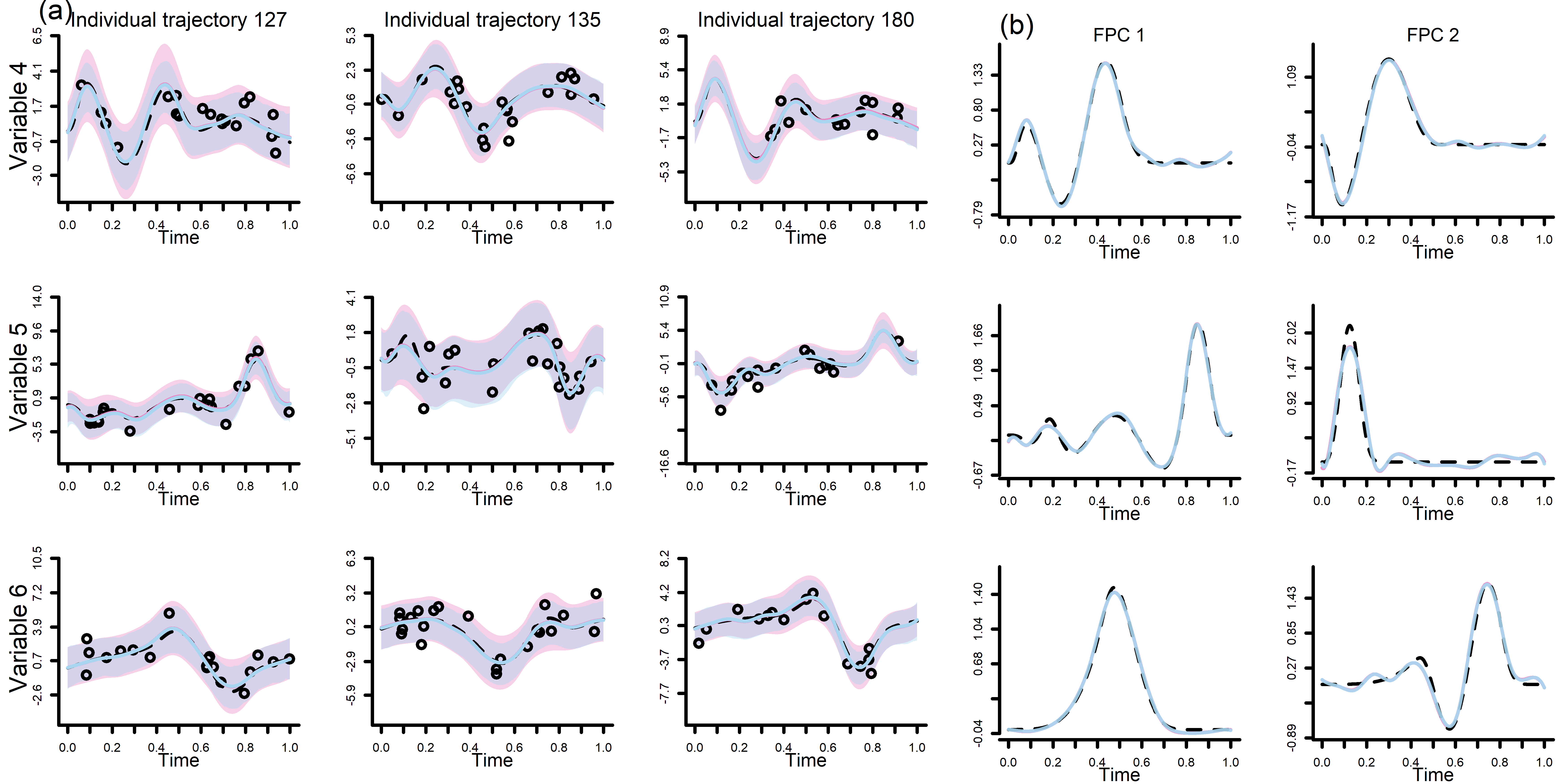}
    \includegraphics[scale=0.4]{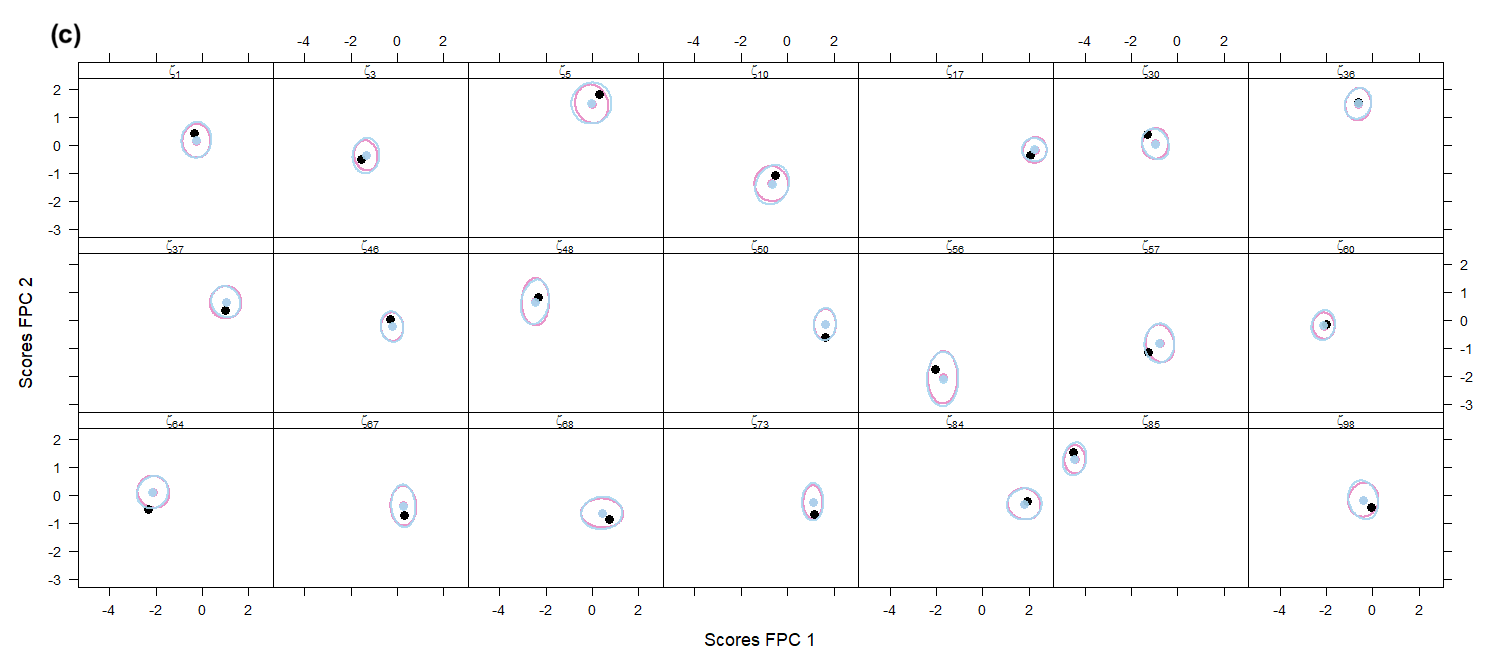}
    \caption{\textbf{Estimated individual curves, eigenfunctions and scores of the PFPCA model obtained from the MCMC algorithm versus our AMFVB algorithm.} Fits obtained from the MCMC (resp. AMFVB) algorithm are in blue (resp. pink).
    (a) The solid lines are the individual curves posterior mean and the colored area the 95\% credibility intervals, for 3 variables in three randomly selected individuals. Observations are displayed as black dots and simulated individual curves are the black dashed lines. 
   (b) The solid (resp. dashed) lines correspond to the 2 estimated (resp. simulated)  eigenfunctions for the 3 same variables. 
    (c) The colored dots are the scores posterior mean and the ellipses are the 95\% credible contours for 21 randomly selected individuals. Black dots are the simulated scores. }
    \label{fig:mcmcVSmfvb}
\end{figure}

Both algorithms successfully recover the true partition, assigning the variables to the group to which they belong with probabilities close to $1$. Figure \ref{fig:mcmcVSmfvb} displays the reconstructed individual curves of the $3$ variables belonging to the same group, for $3$ randomly selected individuals.  
The individual curves estimated via AMFVB and HMC closely match the simulated curves, with highly overlapping $95\%$ credible intervals that capture most of the simulated data points. The simulated eigenfunctions and scores are also well estimated by  both inference methods. Overall, this simulation demonstrates strong agreement in posterior estimation between the two methods.

Inference took less than $2$ minutes using our variational algorithm on a 1-core Intel Xeon CPU, 2.60 GHz. In contrast, the HMC algorithm, run for only $1000$ iterations and $150$ warm-up iterations, required over $14$ hours on the same computer, rendering inference impractical for larger datasets. 

\subsection{Performance on large multi-group problems}\label{subsec:compar}

In this section, we compare the performance of our approach, PFPCA, on larger problems involving structured groups of variables, whereby alternative applications of FPCA may face limitations. 

\subsubsection{Simulation scenarios}
We consider problems with $N=200$ individuals and $p=100$ variables distributed across $Q=4$ groups, with sizes $70$, $20$, $8$ and $2$, respectively. We simulate longitudinal data with $n_\lambda=5$ measurements on average per individual and variable, obtained from $L=3$ eigenfunctions in each of the $4$ groups, as described in Section~\ref{sec_settings}. The variances of the scores are chosen to target the following distribution of proportion of variance across components, within each group $q=1,\dots,Q$: $\text{PV}^{(q)}_1=50\%$, $\text{PV}^{(q)}_2=30\%$ and $\text{PV}^{(q)}_3=20\%$, and calibrated so that the principal components and the error account for $60\%$ and $40\%$ of the total variance respectively, resulting in $v_1=0.83, $ $v_2 = 0.64$ and $v_3 = 0.52$.

\subsubsection{Comparators}
We compare PFPCA with two different applications of FPCA, namely, using:
\begin{itemize}
    \item \emph{univariate FPCA}, applied to each variable independently, i.e., neglecting the shared dynamics underlying each group of variables (hereafter referred to as ``UFPCA''). This model has one set of FPCA scores per variable and per individual;
     
    \item \emph{group-level applications of multivariate FPCA}, i.e., applying  multivariate FPCA \citep{nolan_efficient_2025} independently to each group of variables, assuming that the true partition is known (referred to as ``Oracle MFPCA'').
\end{itemize}
For the existing univariate and multivariate FPCA applications (UFPCA and Oracle MPFCA), we perform inference using the mean-field variational algorithms described in \cite{nolan_bayesian_2025, nolan_efficient_2025}.  
We emphasize that, unlike the above approaches, our PFPCA method does not assume any partition of the variables upfront, but learns it simultaneously with each of the MFPCA expansions of the inferred groups, using joint inference.

To assess the retrieval of the groups, we derive the following \emph{two-step partitioning approach} from the UFPCA estimates, that we use as comparator to our joint PFPCA approach: we obtain posterior estimates of individual scores for $L$ components $\widehat{\bm{\zeta}}_1^{(j)}, \dots, \widehat{\bm{\zeta}}_L^{(j)}$ for each variable $j = 1, \dots,p,$ that we concatenate into a vector of length $L\times N$ denoted $\widehat{\bm{\zeta}}^{(j)}.$ We apply a K-medoids algorithm on these vectors using the distance $$d(\widehat{\bm{\zeta}}^{(j)}, \widehat{\bm{\zeta}}^{(\Tilde{j})})= \min \left\{\sum_{m=1}^{N\times L} \left(\widehat{\zeta}_m^{(j)}+\widehat{\zeta}_m^{(\Tilde{j})}\right)^2; \sum_{m=1}^{N\times L} \left(\widehat{\zeta}_m^{(j)}-\widehat{\zeta}_m^{(\Tilde{j})}\right)^2\right\}, ~1\leq j< \Tilde{j}\leq p.$$  We select the number of partitions $Q^*$ by maximizing the average silhouette over all variables (see Supplementary Material).

\subsubsection{Improving partitioning  using joint inference}\label{subsubsec:simu_partition}

\begin{figure}
    \centering
    \includegraphics[scale=0.7]{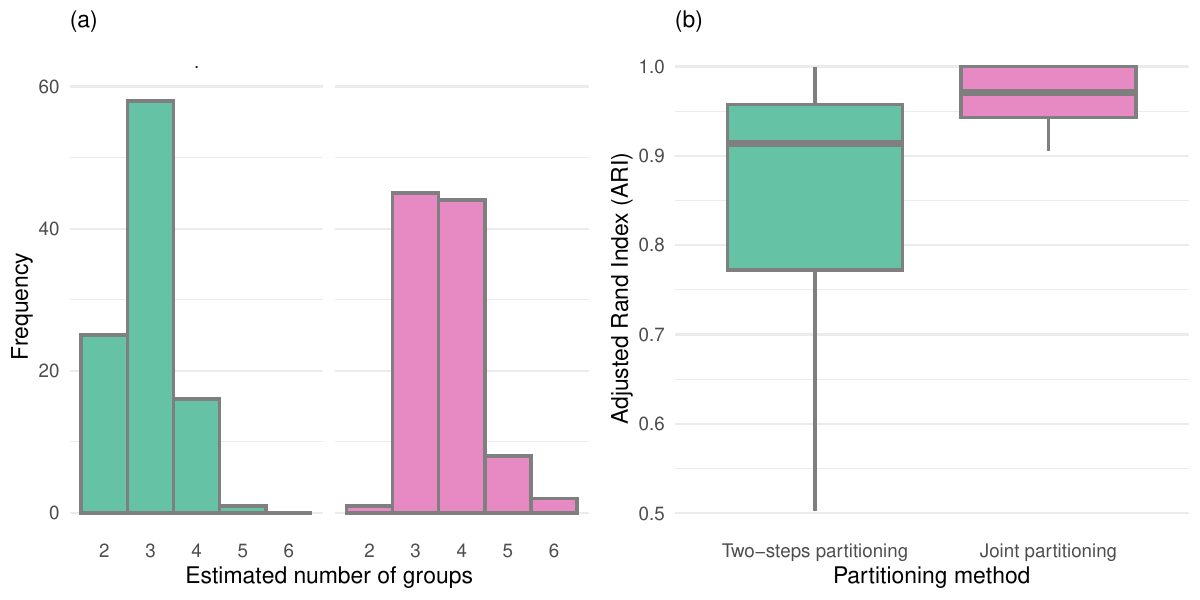}
    \caption{\textbf{Comparison of partitioning estimated by a two-step approach combining UFPCA and a K-medoids clustering, versus PFPCA, for a problem with $Q=4$ groups of sizes $70$, $20$, $8$ and $2$, respectively ($100$ data replicates).} (a) Distribution of number of groups selected; (b) distribution of Adjusted Rand Indexes.}
    \label{fig:ari}
\end{figure}

PFPCA correctly assigns all variables belonging to the three larger groups to their true group in the majority of cases: $93\%$ for the largest group, and $97\%$ and $89\%$ for the second and third groups respectively. In comparison, the two-step approach exactly recovers the first, second and third groups in $13$, $25$, and $25\%$ of the datasets respectively. When including the performances on the smallest group, PFPCA recovers the overall true partition in 27\% of cases and the true number of groups in $44\%$ of cases, as shown in Figure \ref{fig:ari} (a). This performance is primarily limited by this last group, which contains only two variables and is inherently difficult to identify. Indeed, these two variables are assigned to the right group in $34\%$ of the datasets, while in $17\%$ of the datasets, at least one of them is assigned to a singleton, and in $60\%$ of the cases, one or both are absorbed into a larger group. In comparison, the two-step approach only recovers the correct partition in $1\%$ of the datasets, and underestimates the number of groups in $83\%$ of cases, against 46\% with the joint estimation. Figure \ref{fig:ari} (b) displays the distribution of the ARI for the partition estimated with the two-step approach versus PFPCA. This metric is higher in $96\%$ of datasets with PFPCA than when using the two-step approach. In the rare cases (4\%) when the two-step approach outperforms PFPCA, it is because the latter wrongly merges the two largest groups together (Figure \ref{fig:ari}). Overall, PFPCA clearly outperforms the two-step approach, with average (first - third quartiles) $\text{ARI}=0.94 ~( 0.94-1.00)$ and $\text{ARI}=0.85 ~(0.77-0.96)$ respectively.

\subsubsection{Impact of partitioning on estimation accuracy}

\begin{figure}[h]
    \centering
   \includegraphics[scale=0.7]{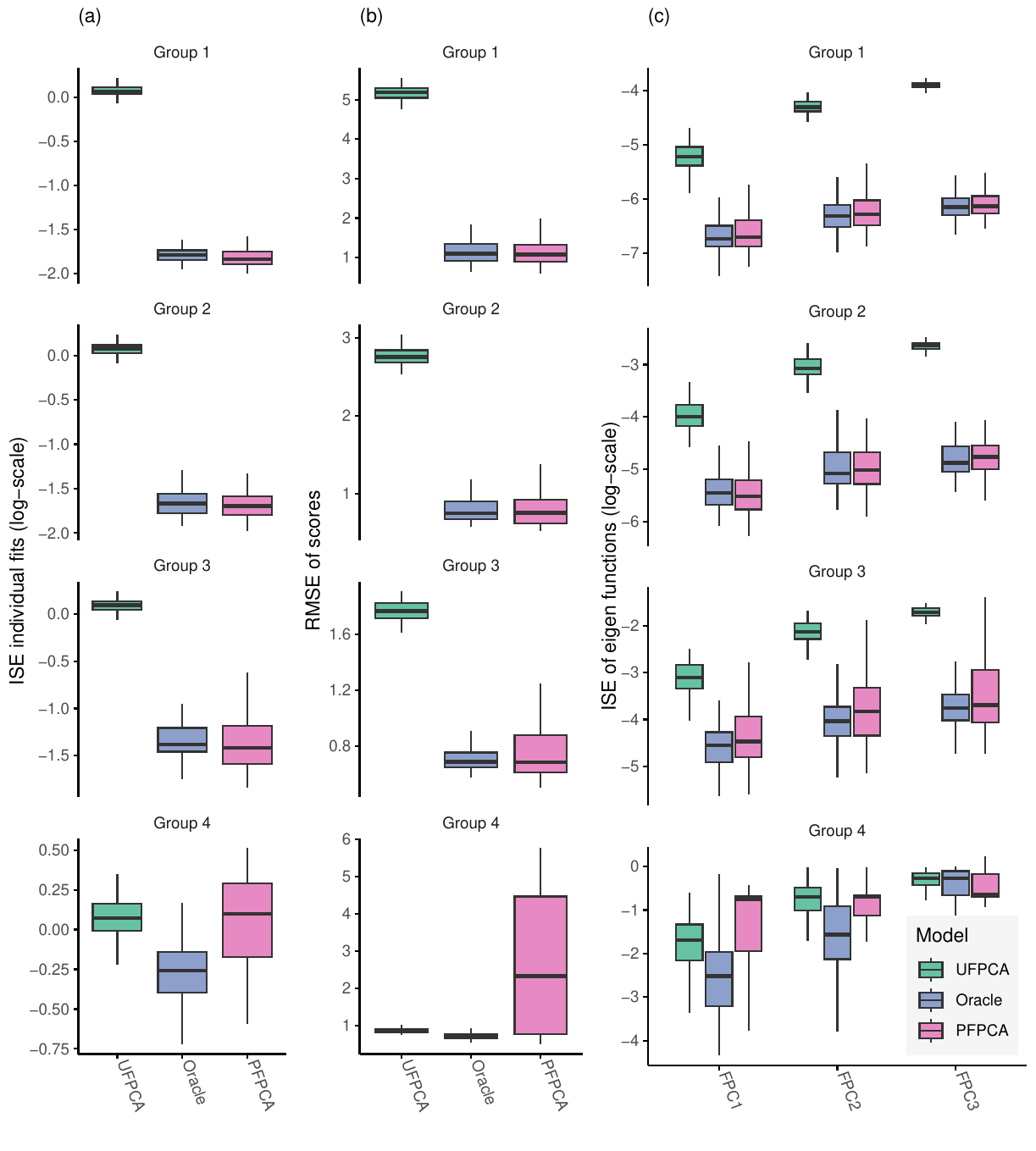}
    \caption{\textbf{Comparison of the functions and scores estimated by UFPCA, Oracle MFPCA and PFPCA for a problem with $Q=4$ groups of sizes $70$, $20$, $8$ and $2$, respectively ($100$ data replicates).} (a) ISE of the individual curves averaged over all individuals and variables; (b) RMSE of the scores; (c) ISE of the eigenfunctions for each FPCA component.}
    \label{fig:summary}
\end{figure}

Figure~\ref{fig:summary} shows accuracy of the reconstructed curves and FPCA scores estimated by all three methods. In particular, for the large groups (groups 1, 2, and to some extent group 3), estimation accuracy under PFPCA is nearly indistinguishable from the Oracle MFPCA approach, indicating that inferring the partition rather than assuming it does not substantially degrade model fit. PFPCA also consistently outperforms UFPCA in these groups, where information borrowing across a large number of variables enables more precise recovery of the functions and scores. The higher errors observed in the smallest group reflect the group misclassification scenarios previously discussed in section \ref{subsubsec:simu_partition} and yield performance comparable to UFPCA when variables are treated effectively as ungrouped. Misclassification mostly impacts the scores, with a large increase in RMSE when variables are absorbed by larger groups, hence  shrinking estimate towards the value of the score in the larger group and inducing bias for the misclassified variable. Overall, the impact of wrong partitioning on estimation accuracy remains moderate.   

\subsubsection{Learning optimal number of FPCA components}

\begin{figure}
    \centering
    \includegraphics[scale=0.8]{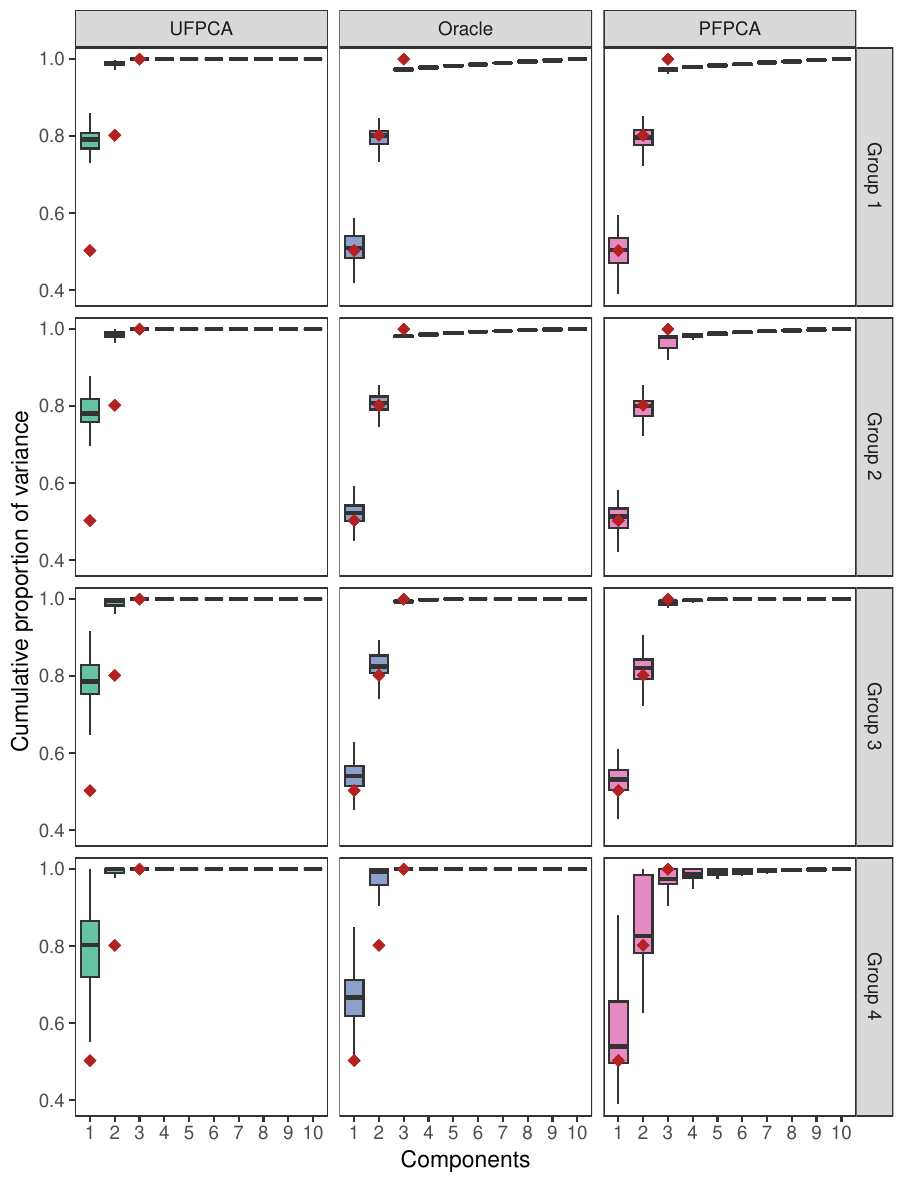}
    \caption{\textbf{Cumulative proportion of variance explained by each component using UFPCA, Oracle MFPCA and PFPCA for a problem with $Q=4$ groups of sizes $70$, $20$, $8$ and $2$ ($100$ data replicates).} Simulated cumulative proportions of variance are displayed as red diamonds. Each column (resp. row) is a different method (resp. group). }
    \label{fig:propvar}
\end{figure}

Figure \ref{fig:propvar} shows that UFPCA systematically overestimates the proportion of variance explained by the first and second principal components, while the Oracle MFPCA and PFPCA approaches accurately estimate the proportion of variance for all three components, in the three larger groups 1, 2 and 3. Similarly to UFPCA, in the smallest group 4, Oracle MFPCA overestimates the proportion of variance explained by the first and second components. In this group, the small number of variables does not bring enough information to identify the variability of the third component, which only accounts for 20\% of the total variability. One could believe that PFPCA is outperforming the Oracle MFPCA in this last group; actually, as we have previously observed, variables of the last group tend to get absorbed by one of the larger groups. As here the simulated distribution of proportion of variance is the same across groups, these variables benefit from borrowing of information from the variables of their estimated group. This does not mean that PFPCA is performing better than Oracle MFPCA, as the distribution of proportion of variance may vary across groups in applications, and therefore wrongly assigning a variable to another group would lead to bias in the estimation of the proportion of variance as well. 

\subsubsection{Varying the density of the observation grid}
To further assess the impact of sparsity of the observation grid on estimation accuracy and on the quality of partitioning, we conduct additional simulations varying the proportion of variables sparsely ($n_\lambda=5$) and densely ($n_\lambda=20$) observed within each group (see Supplementary Material). Increasing the proportion (0, 25, 50, 75 and 100\%) of frequently observed variables leads to a gradual improvement of the partitioning (Supplementary Material), with the average ARI (standard deviation) increasing from $0.94 ~(0.13)$ to $0.98~ (0.08)$ in the rich setting. Correspondingly, the average number of non-empty groups increases from 3.64 (0.76) to 3.99 (0.69) as the amount of data increases, allowing correct identification of variables belonging to the smallest group. Moreover, estimation accuracy improves as the proportion of densely observed variables increases, particularly in smaller groups where information borrowing is limited. In larger groups, however, joint modeling mitigates this effect by leveraging information from the more densely observed variables. Finally, when increasing the density of the observation grid, PFPCA is able to retrieve the correct distribution of proportion of variance in each group (see Supplementary Material).

\section{Evolution of gene expression levels in human influenza }\label{sec:appli}

We now return to the motivating longitudinal gene-expression study of individuals experimentally infected with the H3N2 influenza virus, introduced in Section~\ref{sec-intro}, and use it as a case study to illustrate the practical benefits of PFPCA. With the methodological framework in place, our goals are to (i) estimate groups of genes with coordinated time evolution, (ii) identify, among these groups, the one(s) involved in the response to infection and (iii) emphasize potential associations between the temporal dynamics of these genes and clinical variables, such as presence of symptoms. 

\subsection{Data}

The dataset includes 17 healthy individuals inoculated with H3N2 influenza virus. Blood samples were collected at the same interval times across individuals, with a first measurement the day before inoculation, one sampled shortly after inoculation and 14 additional measurements during the 4 days following inoculation. The samples were assayed by DNA microarray to produce gene expression values for $11\,961$ genes. The symptomatic status of each individual was also recorded. Among the 17 individuals, 9 showed symptomatic response to the infection and 8 did not. 

\subsection{Analysis strategy }

To focus the analysis on genes whose expression variability is more likely to reflect biologically meaningful regulation rather than baseline expression levels, we consider a subset of $1\,000$ genes with the highest variability of expression level relative to their magnitude, according to the coefficient of variation, widely used in genomic analyses \citep{pelabon_use_2020}. For gene $j=1,\dots, 11\,961$, we define the coefficient of variation as the ratio between the empirical standard deviation and the empirical mean of the vector of all samples  $\bm{y}^{(j)}= \left(\bm{y}^{(j)\intercal}_1, \dots, \bm{y}^{(j)\intercal}_N\right)^{\intercal}$ of length $n_{\bullet}^{(j)}$ as follows: $$\text{CV}^{(j)}=\frac{\hat{\sigma}_{\bm{y}^{(j)}}}{\overline{\bm{y}}^{(j)}},$$ where $\overline{\bm{y}}^{(j)}$ is the empirical mean of $\bm{y}^{(j)}$ and $$\hat{\sigma}_{\bm{y}^{(j)}} = \sqrt{ \frac{1}{n_{\bullet}^{(j)}-1} \left(\bm{y}^{(j)} - \overline{\bm{y}}^{(j)}  \bm{I}_{n_{\bullet}^{(j)}} \right)^\intercal\left(\bm{y}^{(j)} - \overline{\bm{y}}^{(j)} \bm{I}_{n_{\bullet}^{(j)}} \right) }.$$ We apply PFPCA, setting both the maximum number of groups and the maximum number of eigenfunctions to $Q=L=10$. To facilitate biological interpretation, we perform an enrichment analysis using the \texttt{KEGG\_2021\_Human} library in the R package \texttt{enrichR}: for each estimated group, the corresponding genes define the target set, while the background is taken to be the set of all other genes included in the analysis. This choice ensures that enrichment is assessed relatively to the same universe of expressed and pre-filtered genes, rather than against the full genome. The \texttt{enrichR} package accounts for multiplicity of hypotheses testing by using the Benjamini-Hochberg (BH) procedure \citep{benjamini_controlling_1995}.

\subsection{Uncovering group structure}
Group sizes vary between 18 and 320 genes. 
 Statistically significant pathway enrichment is found for two of the estimated groups: the largest group, group 4, is enriched for the ``Thermogenesis'' pathway, while group 7, which contains 103 genes, is enriched for the ``Influenza A'' pathway. These findings suggest coherent functional annotation in these cases. The first FPCA component alone accounts for 72 and 79\% of the total variability respectively and, in these two groups, 85\% of the total variability is explained by combining the first two components. For the remaining groups, no pathway reaches significance at level 5\% after multiple-testing correction, likely reflecting more diffuse or context-specific biological functions rather than the absence of structure. The same significant pathways are obtained by applying PFPCA with $Q=20$ mixture components, supporting the robustness of these findings.

      \subsection{Time course of immune response to viral infection}

\begin{figure}
    \centering
   \includegraphics[scale=0.45]{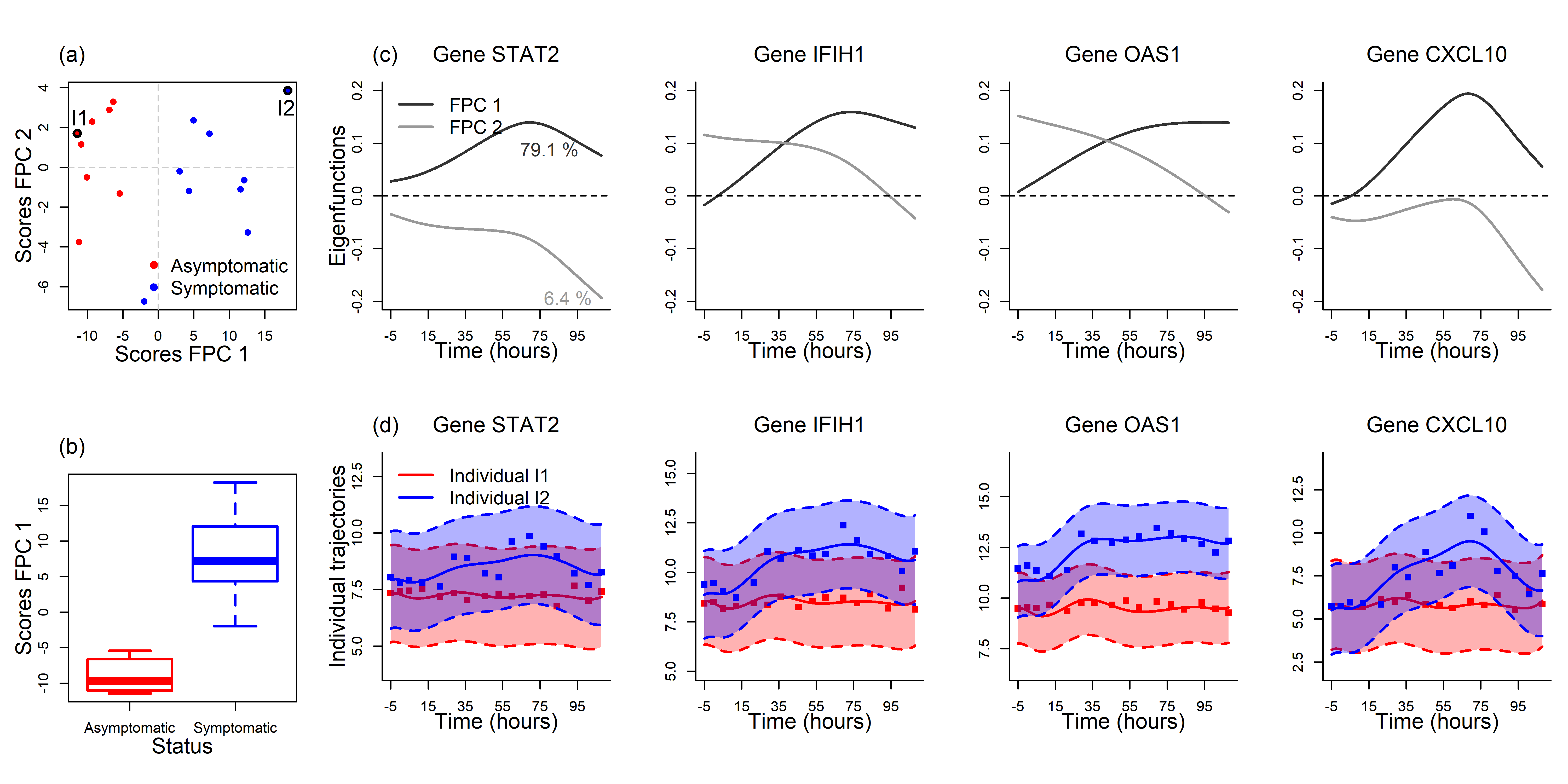}
    \caption{\textbf{PFPCA analysis of a selection of genes belonging to the ``immune response to viral infection'' pathway. }
    (a) Scatter plot of First (FPC1) versus second Principal Component (FPC2) scores, coloured according to symptomatic status. I1 and I2 correspond to individuals with the lowest and highest scores for the first component, respectively. (b) Distribution of first component scores in asymptomatic (red) versus symptomatic individuals (blue). (c) First (black lines) and second (gray lines) eigenfunctions for genes STAT2, IFIH1, OAS1 and CXCL10. (d) Predicted individual trajectories (solid coloured lines), 95\% credible intervals (dashed coloured lines) in individuals I1 and I2, and observed gene levels (coloured squares).    }
    \label{fig:timecourse_immunepathway}
\end{figure}

To explore the immune response to influenza, we examine the ``Influenza A'' group, focusing on a subset of genes known to be involved in the immune response to the infection. As shown  Figure \ref{fig:timecourse_immunepathway} (a)-(b),  
the individual scores associated with the first FPCA component clearly discriminate between symptomatic and asymptomatic individuals. 
This suggests that the individual scores associated with the first FPCA component capture variation related to the symptomatic response to H3N2 infection and may serve as a low-dimensional proxy for this response, potentially useful for prediction purposes. 
     
     Figure \ref{fig:timecourse_immunepathway} (c) indicates that, for genes \emph{STAT2}, \emph{IFIH1}, \emph{OAS1} and \emph{CXCL10}, the first eigenfunction is positive across the time domain. This suggests that higher scores along the leading component correspond to expression trajectories that deviate upwards from the population mean. Consequently, individuals with positive first-component scores exhibit increased expression of these genes relative to the overall cohort. In particular, Figure \ref{fig:timecourse_immunepathway}(d) shows that the individual with the highest score (individual I2) has markedly elevated expression of these influenza pathway genes, consistent with a stronger symptomatic response to infection. In contrast, the individual with the lowest score along the first component (individual I1) shows expression levels that remain close to baseline over time, suggesting limited activation of these immune-response genes and a largely asymptomatic profile. PFPCA was well able to characterize individual temporal evolution of these genes, consistent with goodness-of-fit performances observed in simulations.

     Finally, across all groups except groups 5 and 6, the leading FPCA component captures variation related to the symptomatic response to infection, as indicated by the distinct distribution of individual scores between symptomatic and asymptomatic individuals (see Supplementary Material, Section 6.2). To assess whether the separation between symptomatic and asymptomatic individuals occurs at baseline or emerges later during the host response, we performed a sensitivity analysis in which PFPCA was applied using only the first two observation times: one measurement pre inoculation  and one measurement shortly after inoculation. Interestingly, under this restricted setup, the model no longer distinguished symptomatic from asymptomatic individuals (see Figure 11 in Supplementary Material), suggesting that differences in gene expression evolution become apparent only at later stages of the immune response to infection. This was further confirmed by looking at gene-gene correlations plots at fixed observation times (see Figure 12 in Supplementary Material).

 \section{Discussion}

In this article, we introduced PFPCA, a novel approach for jointly modeling the longitudinal trajectories of many time-dependent variables while grouping them according to shared underlying biological pathways driving their dynamics. We proposed a variational inference algorithm for parameter estimation, enabling substantially faster inference than MCMC-based alternatives such as HMC-NUTS. Through simulations, we showed that joint estimation of all PFPCA parameters can improve the recovery of individual scores, eigenfunctions and partitioning of time-dependent variables, thanks to effective information borrowing. Additionally, our approach is able to learn the optimal number of mixture ($Q$) and FPCA ($L$) components directly from the data, while avoiding time-consuming model selection procedures. We applied PFPCA to a high-dimensional longitudinal transcriptomic dataset of 17 healthy individuals inoculated with H3N2 influenza virus, with repeated measurements of $1\,000$ genes expression level over time. The results demonstrated strong biological relevance and were consistent with previous findings in the literature \citep{chen_predicting_2011, cai_dynamic_2024}.

This work offers many perspectives for future developments. 
Although not encountered in our application, the method already supports irregular time grids across individuals and variables, which is common in many applications, such as longitudinal biomarker monitoring in hospitalized patients. Our framework, however, requires at least one observation point per individual and per variable. Extensions could address missing data mechanisms in which some variables are entirely unobserved for subsets of individuals. More generally, allowing variable-specific time domains could capture delayed or phase-shifted mechanisms driving these variables, which could be useful in other applications, such as modeling the co-evolution of macroeconomics variables. Finally, many biomedical studies involve informative dropout driven by time-to-event outcome such as death. Extending PFPCA to jointly model longitudinal trajectories and survival processes \citep{kerioui_modelling_2022} would be highly relevant, but would require substantial methodological development, given the limited literature on variational inference for survival models.

 %%%%%%%%%%%%%%%%%%%%%%%%%%%%%%%%%%%
\begin{funding}
 This work is supported through the Lopez--Loreta Foundation (M.K., D.T., H.R.). 
\end{funding}

\begin{supplement}
\stitle{Supplementary Material PDF document}
\sdescription{The document provides further details on  1) model specifications; 2) variational updates; 3) sensitivity of the model to prior distributions;  4-5) sensitivity to observation grid density; 6) real-case application.  }
\end{supplement}

\bibliographystyle{imsart-nameyear}  
\bibliography{references}         

\end{document}